\documentclass[prd,showpacs,twocolumn]{revtex4}
\usepackage{amsmath}
\usepackage{amssymb}
\usepackage{times}

\newcommand{\be}{\begin{equation}}
\newcommand{\ee}{\end{equation}}
\newcommand{\const}{\operatorname{const}}
\newcommand{\sgn}{\operatorname{sgn}}

\begin{document}

\title{Uniqueness of the electrostatic solution in Schwarzschild space}

\author{P\'al G. Moln\'ar}
\email{molnar@tp1.ruhr-uni-bochum.de}
\author{Klaus Els\"asser}
\email{els@tp1.ruhr-uni-bochum.de}
\affiliation{Institut f\"ur Theoretische Physik I, Ruhr-Universit\"at Bochum, D-44780 Bochum, Germany\\
	 {\rm (Received 11 October 2002)}}

\begin{abstract}
In this Brief Report we give the proof that the solution of any static test charge 
distribution in Schwarzschild space is unique. In order to give the proof we
derive the first Green's identity written with $p$-forms on (pseudo) Riemannian
manifolds. Moreover, the proof of uniqueness can be shown for either any purely
electric or purely magnetic field configuration. The spacetime geometry is not
crucial for the proof.
\end{abstract}


\maketitle

\section{\label{sec:intro}Introduction}
In 1971, Cohen and Wald \cite{Cohen:1971} presented the electrostatic field of a
point test charge in the Schwarzschild background by using a multipole
expansion. For the same problem, Linet found the solution in algebraic form 
\cite{Linet:1976}. Recently, Moln\'ar generalized the solution for any test 
charge distribution in Schwarzschild space with boundary values 
\cite{Molnar:2001}. He gave the solution not only in algebraic form but also 
in terms of a multipole expansion. Therefore the question
arises whether the 
solution is unique. In this Brief Report, we present the proof that the solution of any
test charge distribution held at rest near a Schwarzschild black hole is unique.

The Brief Report is organized as follows: In Sec.~\ref{sec:green} we derive Green's
first identity written with $p$-forms on (pseudo) Riemannian manifolds. Holmann
and Rummler gave in their book a version for a pair of $p$ and $(p-1)$-forms
on Riemannian manifolds \cite{Holmann:1981}. However, our version is more
general because it is valid for a pair of two $p$-forms and for
pseudo-Riemannian manifolds too. The Schwarzschild spacetime, for example, is
pseudo-Riemannian. In Sec.~\ref{sec:uniq} the uniqueness of the electrostatic
solution is derived with the help of Green's first identity. In 
Sec.~\ref{sec:ae} we replace the spherical symmetry by the axial one and show
that, for the purely electric or purely magnetic solutions, the proof can be
completed along the same lines.

\section{\label{sec:green}The first Green's identity}
Let $(M,g)$ be an $n$-dimensional oriented (pseudo) Riemannian manifold and let
$D$ be a region of $M$ with smooth boundary such that $\bar{D}$ is compact. For
the two $p$-forms $u,v\in\bigwedge_{p}(M)$ we consider the combination
\be \label{eq:2.1}
u\wedge\ast\,d v\in\bigwedge{}_{n-1}(M)\;.
\ee
In the following we use Stokes' theorem, the anti-Leibniz rule and several facts
on $p$-forms which can be found in the literature, e.g., in 
\cite{Holmann:1981,Thirring:1997,Westenholz:1981,Straumann:1984}.
Using Stokes' theorem and the anti-Leibniz rule we obtain
\begin{align} \label{eq:2.2}
\int_{\partial D}u\wedge\ast\,d v &= \int_{D}d (u\wedge\ast\,d v)\notag \\
&= \int_{D}d u\wedge\ast\,d v+(-1)^{p}\int_{D}u\wedge d\ast d v\;.
\end{align}
Now, we write the term $d\ast d v$ in Eq.~(\ref{eq:2.2}) in a different form.
For every form $\omega\in\bigwedge_{k}(M)$
\be \label{eq:2.3}
\ast\ast\omega =(-1)^{k(n-k)}\,{\rm sgn}(g)\,\omega\;.
\ee
This gives us
\be \label{eq:2.4}
(-1)^{k(k-n)}\,{\rm sgn}(g)\,\ast\ast\,\omega =\omega\;.
\ee
With $\omega =d\ast d v$ and $k=n-p$
\begin{eqnarray} \label{eq:2.5}
d\ast d v & = & (-1)^{(n-p)(n-p-n)}\,{\rm sgn}(g)\,\ast\ast\,d\ast d v \nonumber\\
 & = & (-1)^{p(p-n)}\,{\rm sgn}(g)\,\ast\ast\,d\ast d v\;.
\end{eqnarray}
The codifferential $\delta :\bigwedge_{q}(M)\rightarrow\bigwedge_{q-1}(M)$ is
defined by
\be \label{eq:2.6}
\delta := {\rm sgn}(g)\,(-1)^{nq+n}\,\ast d\ast\;.
\ee
We solve Eq.~(\ref{eq:2.6}) for $\ast d\ast$
\be \label{eq:2.7}
\ast d\ast =(-1)^{-nq-n}\,{\rm sgn}(g)\,\delta\;.
\ee
Setting Eq.~(\ref{eq:2.7}) in Eq.~(\ref{eq:2.5}) we obtain for $d\ast d v$ $(q=p+1)$
\be \label{eq:2.8}
d\ast d v = (-1)^{p}\,\ast\delta d v\;.
\ee
We set Eq.~(\ref{eq:2.8}) in Eq.~(\ref{eq:2.2})
\be \label{eq:2.9}
\int_{\partial D}u\wedge\ast\,d v=\int_{D}d u\wedge\ast\,d
v+\int_{D}u\wedge\ast\,\delta d v\;.
\ee
Since for two $p$-forms $\alpha ,\beta$
\[
\alpha\wedge\ast\,\beta =\beta\wedge\ast\,\alpha\;,
\]
Eq.~(\ref{eq:2.9}) becomes
\be \label{eq:2.10}
\int_{\partial D}u\wedge\ast\,d v=\int_{D}du\wedge\ast\,d
v+\int_{D}\delta d v\wedge\ast\,u\;.
\ee

Next, we consider
\[
\delta v\wedge\ast\,u\in\bigwedge{}_{n-1}(M)\;.
\]
Using Stokes' theorem and the anti-Leibniz rule again we have
\be \label{eq:2.11}
\int_{\partial D}\delta v\wedge\ast\,u=\int_{D}d\delta
v\wedge\ast\,u+(-1)^{p-1}\int_{D}\delta v\wedge d\ast\,u\;.
\ee
Now we add Eqs.~(\ref{eq:2.11}) and (\ref{eq:2.10})
\begin{align}
\int_{\partial D}(u\wedge\ast\,d v+\delta v\wedge\ast\,u)
&=\int_Ddu\wedge\ast\,d v +\int_D\Box v\wedge\ast\,u \notag \\
& \quad +(-1)^{p-1}\int_D\delta
v\wedge d\ast u\;. \label{eq:2.12}
\end{align}
This is Green's first identity written with $p$-forms on a (pseudo) Riemannian
manifold, where $\Box := d\circ\delta +\delta\circ d$ is the
Laplace-Beltrami operator.

One can easily derive the second Green's identity \cite{Molnar:2001} with the
help of Eq.~(\ref{eq:2.12}). If we write down Eq.~(\ref{eq:2.12}) again with $u$ and $v$
interchanged, and subtract it from Eq.~(\ref{eq:2.12}), we have
\begin{multline}\label{eq:2.13}
\int_{\partial D}(u\wedge\ast\,d v-v\wedge\ast\,d u+\delta
v\wedge\ast\,u-\delta u\wedge\ast\,v) \\
=\int_D(\Box v\wedge\ast\,u-\Box u\wedge\ast\,v)+(-1)^{p-1} \\
\times\int_D(\delta v\wedge d\ast u-\delta u\wedge d\ast v)\;.
\end{multline}
Note that $d u\wedge\ast\,d v=d v\wedge\ast\,d u$. By definition
(\ref{eq:2.6}) it follows that
\begin{align}
\delta v\wedge d\ast u &= \sgn (g)\,(-1)^{np+n}\,\ast d\ast v\wedge
d\ast u \notag \\
 &= \sgn (g)\,(-1)^{np+n}\,\ast d\ast u\wedge d\ast v=\delta u\wedge d\ast v
 \;. \label{eq:2.14}
\end{align}
With the help of Eq.~(\ref{eq:2.14}) the second term on the right-hand side of
Eq.~(\ref{eq:2.13}) cancels and we find
\begin{multline} \label{eq:2.15}
\int_{\partial D}(u\wedge\ast\,d v-v\wedge\ast\,d u+\delta
v\wedge\ast\,u-\delta u\wedge\ast\,v) \\
=\int_{D}(\Box v\wedge\ast\,u-\Box u\wedge\ast\,v)\;.
\end{multline}
This is Green's second identity which was already derived by Moln\'ar
\cite{Molnar:2001}.

\section{\label{sec:uniq}Uniqueness of the electrostatic solution}
We write Maxwell's equations with the exterior calculus
\be \label{eq:3.1}
d F^{(1)}=0\;,\quad \delta F^{(1)}=4\pi J\;.
\ee
By Poincar\'e's lemma we can introduce a potential form $A^{(1)}=A^{(1)}_\mu\,d x^\mu$
with $F^{(1)}=d A^{(1)}$. Then, the inhomogeneous Maxwell equations become
\be \label{eq:3.2}
\delta d A^{(1)}=4\pi J\;.
\ee
The gauge freedom permits us to choose a special gauge condition for $A^{(1)}$. We
require the Lorenz condition $\delta A^{(1)}=0$. Thus, Eq.~(\ref{eq:3.2}) may be
written in the form
\be \label{eq:3.3}
\Box A^{(1)}=4\pi J\;.
\ee

Now, suppose that there exists another solution $A^{(2)}$ satisfying
the Lorenz condition, Eq.~(\ref{eq:3.3}), and the same boundary conditions. Let
$A\in\bigwedge_{1}(M)$
\be \label{eq:3.4}
A := A^{(2)}-A^{(1)}\;.
\ee
Then
\be \label{eq:3.5}
\Box A=0\,,\quad \delta A=0\,,\quad \left. A\right|_{\partial D}=0\,.
\ee
We set $A=u=v$ in Green's first identity (\ref{eq:2.12})
\begin{multline} \label{eq:3.6}
\int_{\partial D}(A\wedge\ast\,d A+\delta A\wedge\ast\,A)=\int_Dd
A\wedge\ast\,d A \\
+\int_D\Box A\wedge\ast\,A+(-1)^{p-1}\int_D\delta
A\wedge d\ast A\;.
\end{multline}
With the specified properties of $A$, this reduces to
\be \label{eq:3.7}
\int_Dd A\wedge\ast\,d A=0\,.
\ee
In the following we write $F=d A$. $F\in\bigwedge_{2}(M)$ denotes, like $A$,
the difference of two solutions
\be \label{eq:3.8}
F\equiv F^{(2)}-F^{(1)}\;.
\ee
Thus, Eq.~(\ref{eq:3.7}) becomes
\be \label{eq:3.9}
\int_DF\wedge\ast\,F=0\,.
\ee
One can show that
\be \label{eq:3.10}
F\wedge\ast\,F=(F,F)\,\eta\;,
\ee
where
\be \label{eq:3.11}
(F,F)=\frac{1}{2}\,F_{\mu\nu}\,g^{\mu\alpha}g^{\nu\beta}F_{\alpha\beta}
\ee
is the scalar product induced in $\bigwedge_{2}(M)$ and
$\eta\in\bigwedge_{4}(M)$ is the volume element. The uniqueness of solutions of
Eq.\ (\ref{eq:3.3}) can only be shown for particular cases where the scalar
product is semidefinite.

Now, consider the electrostatic potential $\Phi^{(1)}$ of a static test charge
distribution in the Schwarzschild background \cite{Molnar:2001,Linet:1976} and
suppose that there is another solution $\Phi^{(2)}$. Let $\Phi$ be the
difference between $\Phi^{(2)}$ and $\Phi^{(1)}$
\be \label{eq:3.13}
\Phi := \Phi^{(2)}-\Phi^{(1)}\;.
\ee
Then, the vector potential is
\be \label{eq:3.14}
A_\mu=\Phi\,\delta^0_\mu
\ee
and for the Schwarzschild metric we can write
\be \label{eq:3.15}
g^{\mu\nu}=0\quad {\rm for}\quad\mu\neq\nu\;.
\ee
Thus, Eq.~(\ref{eq:3.11}) becomes
\be \label{eq:3.16}
(F,F)=g^{00}g^{ii}(\Phi_{,i})^{2}={\rm definite}\;.
\ee
Since the scalar product (\ref{eq:3.11}) is definite for this special case, we 
can conclude with Eq.~(\ref{eq:3.9}) that
\be \label{eq:3.17}
\Phi_{,i}=0\quad{\rm for\;all}\quad i=1,2,3\;.
\ee
Consequently, inside $D$, $\Phi$ is constant. For Dirichlet boundary conditions
(\ref{eq:3.5}), $\Phi =0$ on $\partial D$ so that, inside $D$,
$\Phi^{(1)}=\Phi^{(2)}$ and the solution is unique.

So far we restricted our analysis to the static case. However, the proof of
uniqueness can be straightforwardly generalized to the case of any purely
electric or purely magnetic field configuration, because one can infer with the
scalar product (\ref{eq:3.11}) that all solutions are unique for which $F_{0i}$
or $F_{ij}$ (i.~e.\ $E_i$ or $B_k$) vanish and the
scalar product of $\mathbf{E}$ or $\mathbf{B}$ is definite.

\section{\label{sec:ae}Stationary and axisymmetric systems}
We call a system stationary and axisymmetric when all physical quantities,
including the metric tensor components $g_{\mu\nu}$, which describe the system
are independent of time $t$ and of a toroidal angle $\varphi$. We choose a
coordinate system $(x^\mu )$ with $x^0=ct$ $(c=1)$, $x^1=\varphi$, and $x^2$,
$x^3$ some poloidal coordinates. Assuming in addition that all physical
quantities are invariant to the simultaneous inversion of $t$ and $\varphi$ ---
which is reasonable for any rotating equilibrium --- the most general line
element $ds$ can be represented as follows \cite{Chandra:1992}:
\begin{gather} \label{allequations}
\begin{split}
(ds)^2=g_{rs}\,d x^rdx^s+g_{ab}\,dx^adx^b\;,\\
g_{ab}=0\quad\text{for}\quad a\neq b\;,\
\end{split}
\end{gather}
where the indices $r,s$ run from $0$ to $1$, and $a,b$ from $2$ to $3$. Now, it
is useful to choose particular poloidal coordinates, as defined by the poloidal
stream lines, $\Psi =\const$, and an angle-like coordinate $\theta$ varying
along the poloidal stream lines
\be \label{eq:4.2neu}
x^2=\Psi\,,\quad x^3=\theta\,.
\ee
Then, we denote the projections of the $j^\mu$ lines onto the poloidal plane as
the lines $\Psi =\const$ and the stream function $\sim I$ of $j^a$ is denoted as
a flux function, $I=I(\Psi )$. The continuity equation for $j^\mu$ in the
poloidal plane is then solved as follows:
\be \label{eq:4.2nneu}
j^2=0\,,\quad 4\pi\sqrt{-g}\,j^3=I'(\Psi )\,,
\ee
where the prime of $I$ means differentiation with respect to $\Psi$.
Els\"asser \cite{Elsasser:2000} showed in his paper that Amp\`ere's equation in
the poloidal plane becomes
\be \label{eq:4.2}
\sqrt{-\frac{g_{\rm sym}}{g_{\rm pol}}}\,F_{23}=I(\Psi)\;,
\ee
where
\[
g_{\rm pol}\equiv{\rm det}(g_{ab})\,,\quad g_{\rm sym}\equiv{\rm det}(g_{rs})\,.
\]
For the difference of two solutions (cf.~Eq.~(\ref{eq:3.8})) we obtain
\be \label{eq:4.3}
F_{rs}=\partial_rA_s-\partial_sA_r=0\,,\quad F_{ab}=0\,,\quad F_{ar}=A_{r,a}\,.
\ee
Hence, the scalar product (\ref{eq:3.11}) gives us
\vspace{0.1em}
\onecolumngrid
\hspace{-1em}\rule{24.5em}{0.05em}\rule{0.05em}{0.8em}
\be 
\label{eq:4.4}
(F,F)=\underbrace{A_{0,a}\,A_{0,b}\,g^{ab}}_{<0}\underbrace{g^{00}}_{>0}
+\underbrace{A_{1,a}\,A_{1,b}\,g^{ab}}_{<0}\underbrace{g^{11}}_{<0}
+2A_{0,a}\,A_{1,b}\,g^{ab}g^{01}\,.
\ee
We see again that uniqueness is only obtained in general if either $A_0$ or
$A_i$ are zero, i.~e.~, if either the electrostatic field or the magnetic field
vanishes. This is also true for a diagonal metric (Schwarzschild, Minkowski, 
$g^{01}=0$). In other words, the symmetry of the spacetime geometry is not
crucial for the proof.
\twocolumngrid

%\bibliography{literatur}
%\bibliographystyle{apsrev}

\end{document}